# On a mathematical frame of reference

Aasis Vinayak P G [$1]

**Abstract**

This papers aims at revisiting Minkowski space-time with a modified outlook and making it more consistent (III.8). The paper scrutinizes the special case of Einstein's relativistic hypothesis ('STR'). The paper tries to solve the problems faced by relativistic hypothesis by proposing a *purely mathematically formulated* reference (and *not* a frame like aether) frame Ω that suits string theory. Then checking the validity of Ω by applying to various phenomenon and aspects explained by 'STR'; and also Ω's properties when the experimental statistics are taken into account in a *homogenous and isotropic large-scale structure* of Universe. Further more the paper checks whether introducing Ω solves the problems in 'STR'; then looks into the occurrences of new paradoxes in Ω, if any present.

[#]References to *basic* equations are not given; as they are easily recognizable - Also if a symbol is left with not *tags indicating what it is;* it will have usual meaning.

## I. Introduction

In special relativity, there are various inconsistencies that cannot be ruled out *as mere paradoxes*. One such issue is that when an observer in a frame moves with a velocity greater than c (which is allowed in relativistic Quantum Mechanics [+1]), the velocity of light is constant for him as per 'STR'[1]. Thus for him a photon which is moving parallel to this velocity can hit a projected zone of the frame, while the same event will be missing for an observer who is in a 'relatively' rest frame since for that event to happen

---

[$1] Department of Physics, Loyola College, Chennai, India 600034
[$1] Varadakshina, Kollam, Kerala, India –691503
Electronic mail – aasisvinayak@gmail.com

[+1] When speaking about wave velocity of tardons it will have velocity greater than c as the product of wave velocity and particle velocity is a constant, i.e. $c^2$ [#].

for him the velocity of light should be greater than c which is prohibited is 'STR'. Thus there is missing of a whole event. Another illustration is the one in which an observer in an entangled two photon system, can easily detect whether he is motion by looking at the other photon. Thus he may find that inertial frame to be moving[+1.a.]. The problem becomes more aggravated when we treat photon as a particle. We cannot answer why *this particle's* velocity is constant and why not other tardonic particles. The fact that a photon is having *zero rest mass* cannot account it logically. Apart from these there are many papers [2] [3] [4] [5] [6] [7] [8] [9] that projects problems with constancy of light and also about the second postulate.

When we contemplate over here, we may go back and check with the postulates of relativity. Soon one finds that there is nothing wrong in the first postulate that helps us to apply laws in their same form and not against our logic. But when it comes to the second one, one may feel bit uneasy yet becomes pacified once he sees that it can Maxwell's equations *invariant in form* and *STR* can explain so many phenomenon. Even if one tries to review it, one cannot go back to the aether based elucidations. There are ample reasons to abandon the ether theory. Thus comes to conclusion that *if at all* anything can replace this, then the same should detail (1) *all phenomenon*[#] elucidated by 'STR' and should have the (2) *same* 'Lorentz transformation' [10] [10.a] but may be by another mode of approach; and thus (3) should leave the Maxwell's equations[#] in its *own form*.

# II. Formulation

## 1.A. Examination of Space-time

Minkowski space seem to have a well defined structure as it is done by means of mathematics with less assumptions. This formulation is accepted here also. But outlook changes are described in section 2.

Before pursuing further here are the assumptions we make for the formulation

  i.  Space-time is treated in four dimensional and hence the allied theorems are used
  ii. A purely mathematical frame $\Omega$ analogous to the celestial sphere (for purpose and not in shape) that we deal in astronomy. $\Omega$ is that frame where the velocity of light is c. All measurements will be based on $\Omega$.
  iii. All laws of physics will retain its form in all frames of references.

---

[+1.a.] Consider an observer in a photon A. He is *aware* of the fact that there is an *entangled* photon B such that 'B' will move only in line parallel to A with the same velocity. But for time being he is not is a position to *see* as there is a *screen* between them. Once the *screen* is removed he finds that B is in motion due to the consequence of second postulate of 'STR'. Thus he could easily conclude that his inertial frame is also moving. We can understand that the awareness *will not come* to play much to help 'STR' as one has the *same awareness* when we analyze the situation when an observer positioned in a body is looking at another body, which is attached to it using a rigid rod, *cannot* detect the similar notion.

To have (2), we *cannot* follow the steps *exactly* as in Lorentz own paper or the *steps* of Einstein. To do that, let there be two frames S and S'. And there is nothing wrong in having a purely mathematical frame, $\Omega$ (as we don't need any *physical* aspect of the frame like how it *helps* in propagation of light *and all* - except the fact that $\Omega$ is the frame in which velocity of light is taken to be velocity of light in vacuum c). We assume that S is at rest with respect to (wrt) $\Omega$ and S' is moving with a velocity v wrt $\Omega$; along x-axis of S and $\Omega$.

Let at time t=0, and at time t=t in $\Omega$ we take *snaps* (a still *photograph*) of the frames (to have *static co-ordinates*) and attach it (*mark co-ordinates*) to $\Omega$.

If X' is the axis on S' and X the axis on S; for the co-ordinates[#] we will have [+2]
$$x' = Ax + By + Cz + Dt + E \quad (1.1)$$

For the motion along X-axis let us assume for *simplicity* that the point under consideration is independent of y & z; and at t=0 in $\Omega$, we have
$$E = 0 \quad (1.2)$$
At time t=t in $\Omega$, for the origin in S' we have
$$0 = Avt + Ht \quad (1.3)$$
Thus
$$x' = A(x - vt) \quad (1.4)$$
Replacing letter 'A' with '$\xi$' (to avoid *confusion*) we obtain
$$x' = \xi(x - vt) \quad (1.5)$$
The inverse *outlook* will fetch
$$x = \xi(x' + vt') \quad (1.6)$$
Consider *one point* just as we considered for (1.2) and (1.3) such that [+3]
$$x' = ct' \quad (1.7)$$
It should be noted that it could be only for one point [also from (1.A) & (1.B)]; others it may not be in this form (say for mentioning two co-ordinates *at same time* see section 2). If so we can have
$$x = ct \quad (1.8)$$
Thus (1.7), (1.8) will be the corresponding point. Applying to (1.5) & (1.6) we get
$$ct = \xi t'(c + v) \quad (1.9)$$

---

[+2] It should be noted that the steps from (1.1) to (1.8) is similar to standard Lorentz transformation equations derivations but for (1.7) and (1.8) the case considered is different

[+3] It should be noted that we are now taking the '*snap*' of the frames and we are *considering the static marked co-ordinates* in $\Omega$.

As we are dealing in Minkowski space[#], for these static points and plotted in $\Omega$, and c is the velocity of light in $\Omega$ we have
$$s^2 = c^2 t^2 - x^2 \quad (1.a) \text{ (For S), if so we get}$$
$$s^2 = c^2 t'^2 - x'^2 \quad (1.bB)$$
[(1.bB) For S' frame measured in $\Omega$ and the given one is not s' when measured in moving frame]

$$ct' = \xi t(c-v) \quad (1.10)$$
$$c^2 tt' = \xi^2 tt'(c^2 - v^2) \quad (1.11)$$
$$c^2 = \xi^2 (c^2 - v^2) \quad (1.12)$$
$$\xi^2 = \frac{c^2}{(c^2 - v^2)} = \frac{1}{1 - \frac{v^2}{c^2}} \quad (1.13)$$
$$\xi = \frac{1}{\left(1 - \frac{v^2}{c^2}\right)^{\frac{1}{2}}} \quad (1.14)$$

Thus we get

$$x' = \xi(x - vt) \quad (1.5)$$
$$y' = y \quad (1.15)$$
$$z' = z \quad (1.16)$$
$$t' = \xi\left(t - \frac{v}{c}x\right) \quad (1.17)$$

Hither forth $\xi$ is used for any $\xi(v)$, where v is the velocity associated.

## 1.B. Deductions

Thus we got akin relations to that of Lorentz transformations[#][10][10.a] as required. Now it is necessary to detail about the differences. Here it is free from the concept of aether as well as the second postulate of special case of relativistic hypothesis of Einstein as stated above. Also using $\Omega$ the velocity of light can be made to change under transformations as viewed from different frames (as said in section 3). The following properties of $\Omega$ make the formulation distinct from Lorentz's *aethero-centric theory*[#][10].

## 2. Properties of $\Omega$

Here if one needs to have two points at same time t, they could be $ct_\alpha$ and $ct_\beta$ and these two times are different from t in $\Omega$ [as mentioned in above description esp. in section (1.A)] .As the final results are akin to Lorentz transformation it can be proved by simple mathematics that it will fetch the length contraction and time dilation as the same in 'STR'.

Now it is necessary to look at the model of $\Omega$. As the universe is homogeneous and isotropic on large scale the *three-dimensional* space (2-sphere in 3 dimension for *space alone*) of $\Omega$ will be spherical. If it so, it should be one which pass through the *points* of consideration and *direction* which are separated by small distance and *appear like a plane*; for that the radius of the sphere should be appreciably large. For another set of points the sphere drawn may be different. Further more the *additional dimension* for time can be *fitted* into this. This may seem like the 3-sphere in 4 dimensions. Thus when viewed from large scale perceptive $\Omega$ will appear like array of points distributed at small distance homogenously and isotropically in the universe. (For a small displacement the *matching* sphere for the *point and direction* remains almost unaltered). Moreover $\Omega$ is the *absolute model* of Minkowski space-time and that in S' is only a transformed one in $\Omega$.

## 3. Composition of Velocity - Velocity Addition

If $\kappa$ be the time in $\Omega$ and if two frames moves at velocities v and u wrt to $\Omega$ along its X-axis in opposite directions, we have for the distance X in $\Omega$,
$$x = (v+u)\kappa \quad (3.1)$$
$$\frac{x}{\kappa} = (v+u) = \Xi \quad (3.2)$$
For the change of co-ordinates we have (from x' to x)
$$x' = \xi(x-vt) \quad (3.3)$$
And for transformed length X' we have
$$x' = \frac{x}{\xi} \quad (3.4)$$
$$\kappa' = \xi \times \kappa \quad (3.5)$$
So the velocity wrt to one of them is given by
$$\Xi' = \frac{x'}{\kappa'} = \frac{(v+u)}{\xi^2} \quad (3.6)$$
$$\Xi' = \frac{|v-u|}{\xi^2} \quad (3.7)$$
This is the *new* result [see section 4 to have meanings of (3.6) & (3.7)].
Now in if one frame is moving is the axis y (with velocity v along y-axis wrt $\Omega$) and to find the velocity wrt to the one moving in x-axis (say for one with velocity u along x-axis wrt $\Omega$), we have (here the time in $\Omega$ is $t_0$)
$$L_0 = \sqrt{x^2 + y^2} = t_0\sqrt{u^2 + v^2} \quad (3.8)$$
$$x' = \frac{L_0 \cos\theta}{\xi} \quad (3.9)$$
$$y' = L_0 \sin\theta \quad (3.10)$$
$$t' = \xi t_0 \quad (3.11)$$
$$v_x' = \frac{L_0 \cos\theta}{\xi^2 t_0} \quad (3.12)$$
$$v_y' = \frac{L_0 \sin\theta}{\xi t_0} \quad (3.13)$$
$$v' = \sqrt{\left[\frac{L_0^2 \cos^2\theta}{\xi^4 t_0^2} + \frac{L_0^2 \sin^2\theta}{\xi^2 t_0^2}\right]} = \sqrt{\frac{L_0^2 \cos^2\theta + \xi^2 L_0^2 \sin^2\theta}{\xi^4 t_0^2}} \quad (3.14)$$

$$v' = \sqrt{\frac{t_0^2(u^2+v^2)\cos^2\theta + \xi^2 t_0^2(u^2+v^2)\sin^2\theta}{\xi^4 t_0^2}} = \frac{1}{\xi^2}\sqrt{(u^2+v^2)\cos^2\theta + \xi^2(u^2+v^2)\sin^2\theta} \quad (3.15)$$

## 4. Interpretation of equation

Equations (3.6) & (3.7) do not tell us that if we measure the velocity of light, in any 'forward' and 'backward' direction by remaining in touch with a frame – say being in earth's surface and measuring the velocity in *forward* and *backward* directions horizontally-[+4]. But rather they are equal as here if we use to find velocity by dividing a specific distance, $L_0$ in time $t_0$ when transported to $\Omega$ will give same transformed results for these two cases as here direction will not come to play.

But equation (3.15) tells us that some thing *similar* that could be seen when it is for 'horizontal' and 'perpendicular'. But to find the difference as predicted by the equation the *change in distance* (*points and direction*) should be small as there is chance for changing the reference sphere when dealing in large scale (for small distance the difference tends to zero). As all the objects in universe are moving away with the velocity there is more chance for *sphere skipping* so this problem will exist if we try to measure from earth.

Now let us see why the velocity addition equation in STR[#] is invalid here, for that we have

$$u_x = \frac{u_x' + v}{1 + \frac{u_x' v}{c^2}} \quad (4.1)$$

But in $\Omega$ there could be points for which
$$dx = c\,dt \quad (4.2)$$
$$dx' = c\,dt' \quad (4.3)$$

Thus,

$$\frac{dx}{dt} = c = \frac{\frac{dx'}{dt'} + v}{1 + \frac{dx'}{dt'}\frac{v}{c^2}} = \left[\frac{\xi^2 c + v}{c + \xi^2 v}\right] \times c \quad (4.4)$$

This is true only if v=c, hence (4.1) is not a generalized equation and cannot be used.

## 5. Electromagnetism

If c is the velocity of light in Ω, for a frame S' moving with a velocity v wrt to Ω will have the Maxwell's equations in the following forms. For the Maxwell's tensor (stress)[#] as viewed from the S' frame looks using (1.5) (1.16) (1.17) (1.18) – (as transformations are akin to that of Lorentz we can have new equations as in the same form as predicted and is like we do problems in classical electrodynamics [11] using 'STR' since *constancy of speed of light* as predicted by 'STR' has got nothing to do here[1])-

$$\begin{pmatrix} \varepsilon_0' E_x'^2 + \dfrac{B_x'^2}{\mu_0'} - \zeta'_{EM} & \varepsilon_0' E_x'^2 E_y'^2 + \dfrac{B_x'^2 B_y'^2}{\mu_0'} & 0 \\ \varepsilon_0' E_x'^2 E_y'^2 + \dfrac{B_x'^2 B_y'^2}{\mu_0'} & \varepsilon_0' E_y'^2 + \dfrac{B_y'^2}{\mu_0'} - \zeta'_{EM} & 0 \\ 0 & 0 & -\zeta'_{EM} \end{pmatrix} \quad (5.1)$$

$$\text{Here } \zeta'_{EM} = \dfrac{\varepsilon_0' E'^2}{2} + \dfrac{B'^2}{2\mu_0'} \quad (5.2)$$

As in S' one has got very freedom to think of the change as all the quantities[11] [now *for the time being –till (5.5) and (5.6):* leave $\mu_0$ and $\varepsilon_0$ ] that can be subjected to Lorentz transformation as we did for *few* in the case of classical electrodynamics[11] [#] as

$$E'^* = \xi(E + \beta \times B) - \dfrac{\xi^2}{\xi+1}\beta(\beta \bullet E) \quad (5.3)$$

$$B'^* = \xi(B - \beta \times E) - \dfrac{\xi^2}{\xi+1}\beta(\beta \bullet B) \quad (5.4)$$

And for similar variations in $\mu_0$ & $\varepsilon_0$ one can resort to reported predictions[12]. Hence the factors mentioned in (5.1) and (5.2) may vary from (5.3) and (5.4) as predicted as by[+5]
For the similar variation for electromagnetic tensor we can have

$$B_x' = \dfrac{\partial A_z'}{\partial y'} - \dfrac{\partial A_y'}{\partial z'} \text{ And so on} \quad (5.5)$$

$$E_x' = -\dfrac{\partial \varphi'}{\partial x'} - \dfrac{\partial A_x'}{\partial t'} \text{ And so on} \quad (5.6)$$

Thus for the electromagnetic tensor[#] (*due the same reasons as stated above*) as viewed from the S' frame looks as

---

[+4] In this case the observation normally[13] made will be by referring to a distance traveled by light wrt to the given frame. But if one does with out this reference it will record a change (for those directions)

$$\begin{pmatrix} 0 & \dfrac{-E_x'}{c'} & \dfrac{-E_y'}{c'} & \dfrac{-E_z'}{c'} \\ \dfrac{E_x'}{c'} & 0 & -B_z' & B_y' \\ \dfrac{E_y'}{c'} & B_z' & 0 & -B_x' \\ \dfrac{E_z'}{c'} & -B_y' & B_x' & 0 \end{pmatrix} \quad (5.7)$$

There is ample reason to believe (plausibly)

$$c' = \dfrac{1}{\sqrt{\varepsilon_0' \mu_0'}} \quad (5.8)$$

For the changes in $\varepsilon_0$ and $\mu_0$ to (5.8) it requires similar conditions and nature as detailed in papers concerned [2].

## 6. Reanalyzing the problems and search for new ones

From the above description, it is unambiguous that the introduction of Ω can solve the entire problems detailed above (as the velocity of light is not a constant here *as in* 'STR'). To suggest a *new paradox* it can be thought that when two photons are moving in opposite directions along the X-axis of Ω, if an observer in one photon looks at another it will appear to be static. But on further analysis we can find that this not a 'paradox' as this effect is due to the contraction of the length between them in Ω as viewed from one photon (just like we have Lorentz length contraction which cannot be called as a 'paradox' as it is not against one's logic when viewed from a mathematical point of view).

## 7. Remarks

In experiments like Michelson-Morley experiment also[13], the velocity of light is appeared as constant in horizontal directions, but when it comes to the vertical direction there could be a change (which they neglected owing to experimental error, but need not be so [14]) partially due to (3.15) [as earth could be also moving is the direction along which the system of the universe is moving – apart from its revolution; hence the sphere skipping may come to play since the original experiment is done on 'large scale' in Ω frame.

Yet it attention-grabbing to note two papers that predicted an absolute frame (but they referred to as aether wind with *its properties*) by analyzing the data - one is 1902 [14]; and then in 2005 [15] which included modern experiments also. This can explained with the purely mathematical Ω.

Another aspect arises when we consider the data from cosmology that our universe is expanding at an accelerated rate. So the velocities of all objects are subjected to change on large scale. As per the formulated Ω, the velocity of light (wrt to a frame) should also vary in that respect. This is in accordance with the data that the velocity of light changed as during the present state of cosmos is evolved [16].

It is also interesting to note that there is an argument similar to - "*One possible solution to the problem of time in quantum mechanics (and thus in quantum gravity) is the reintroduction of a background Newtonian time*" [17] [18]. The above two papers aims to have only an absolute-frame and not *exactly aether* (though aether is not encouraged due logical reasons); both specify some qualities of that space-time.

In fact Ω satisfies all the qualities as required by them (thus Quantum Gravity). There is another paper [19] that also has similar notions yet the space-time is not clear. With the *introduction of Ω, the factors* listed above could be elucidated lucidly.

The further impacts are there could be an absolute space-time scale in Ω, thus the concepts in Einstein's Relativistic Hypothesis needs to get with-drawn to the level of speaking *'some thing which is with respect to (wrt) another'* and the factors in Ω are good to handle effectively than speaking something in *any* 'wrt language'. Also on contemplation we find Eddington's religious fervor is against logic in context of Ω

Moreover it becomes easy to explain the mechanics of accelerated frame (as detailed in "General theory of Relativity") by extending this Ω to that case. (Instead of having a curved space-time we can go for *sphere skipping* with *positive motivation*. But here also 'GTR' will be inadequate one, for solving this we can used the concept of Ω again to explain factors (See section III.8. also). This becomes more relevant when we look at reported data [20] and propositions [21]. More Ω being a mathematical frame can replace (and solve the defects) in the papers (which are having various *other inconsistencies* – see *section 9*) like 'aether theory' [22]

Another implication comes from the fact that Ω can be extended to explain any other results [23] [24] [25] [26] [27] that 'STR' cannot. The relevance of Ω becomes more strong when consider analytical papers on [28] 'STR'. Ω could solve all the problems detailed (including about *asymmetries of Maxwell's electrodynamics*) in Einstein's paper[1].

An analysis of Ω shows that space-time when viewed from the point of view of Ω is not continuous but discrete. This is due to the fact that the centers of the sphere's drawn should be at least at plank's length. Thus when we view the final 4-dimentional structure there will be small 'spaces' where the laws of Ω hold water. This implies that space will have meaning only if it is viewed from this discrete mode. Further more Ω gives ample support to background dependent theories like the string theory, thus giving meaning to the aspects covered by these theories.

## Differences with STR and analysis

With regard to getting results for Lorentz contraction[10] and time dilation[10.a] will have same results; and phenomenon like drag effect (III.9) $\Omega$ will have similar deductions as of STR. But when it comes to the velocity addition theorem and measuring the velocity of light measured by different frames of references with different velocities with respect to $\Omega$ will be different (II.3)

On further analysis it is found that the major problem of STR lies with its second postulate as in $\Omega$ we have made no assumption with regard to the constancy of light yet proved that it cannot be so. More over all the three assumptions are consistent with the established laws of physics.

# III. Supplementary Section

All symbols used below are having their *usual meaning*

**8. General**

For tensor analysis[#] in any S' we can have *(using the reason we used for section 5)*

$$x_\alpha' = (c't, -\varsigma') \quad (8.1)$$
$$x^{\alpha'} = (c't, \varsigma') \quad (8.2)$$
$$x_\alpha' = g_{\alpha\beta} x_\beta' \quad (8.3)$$
$$s'^2 = g_{\alpha\beta} x^{\alpha'} x^{\beta'} \quad (8.4)$$

$$g_{\alpha\beta} = \begin{pmatrix} 1 & 0 & 0 & 0 \\ 0 & -1 & 0 & 0 \\ 0 & 0 & -1 & 0 \\ 0 & 0 & 0 & -1 \end{pmatrix} \quad (8.5)$$

This will also be same. But
$$s \neq s' \quad (8.6)^{+4}$$
So Poincare general transformations[#] that follow (8.6) are *invalid* (but both have *Minkowski space* as said in section 2).

Also for electromagnetism[#] we can have *(using the logic we used for section 5)*

---

[+5] As we have
$$E'^* \neq E' \quad (5.a)$$
$$B'^* \neq B \quad (5.b)$$

$$\nabla' \Box A' + \frac{1}{c'^2} \frac{\partial \phi'}{\partial t'} = 0 \quad (8.7)$$

$$A' = (\frac{\phi'}{c'}, \overrightarrow{A'}) \quad (8.8)$$

$$\partial \varphi' = \partial'_\alpha (\phi') dx'^\alpha \quad (8.9)$$

$$\partial'^\alpha = \left( \frac{1}{c'^2} \frac{\partial}{\partial t'}, \nabla' \right) \quad (8.10)$$

$$\partial'_\alpha = \left( \frac{1}{c'^2} \frac{\partial}{\partial t'}, -\nabla' \right) \quad (8.11)$$

$$\Box' = \partial'_\alpha \partial'^\alpha = \frac{1}{c'^2} \frac{\partial^2}{\partial t'^2} - \nabla'^2 \quad (8.12)$$

$$\Box' A'^\alpha = 0 \quad (8.13)$$

Thus jumping to (5.8) (sensibly)

$$c' = \frac{1}{\sqrt{\varepsilon_0' \mu_0'}} \quad (8.14)$$

## 9. Relativistic Optics

For *drag effect* a simpler method (which differ in *concepts –as not aether is there in* Ω but steps being similar) of the *Frenel's method of deduction using aether*[29] can be applied. Similar steps[29] without aether or steps similar to that of Einstein[1], as he used only transformation equations (and his velocity composition will not come to play), could be used in the case with *Doppler effect* (with regard Ω)[#] and *aberration*[#] (as the transformation equations are akin to Lorentz transformation[#]) and both could also be handled effectively with Ω. But these factors cannot be explained using aether dependent (for aberration using aether dragged by earth) frames [30] [31] [32]. For radiation pressure[#] also we can use steps like the ones in section 5.

## 10. Four Velocity and four acceleration (in Ω)

The formats of the these quantities[#] will also remain unaltered as shown below (As still we have Minkowski space in Ω)

$$d\tau^2 = \frac{ds^2}{c^2} = dt^2 - \frac{dx^2 + dy^2 + dz^2}{c^2} \quad (10.1)$$

$$\frac{d\tau^2}{dt^2} = \left( 1 - \frac{u^2}{c^2} \right) \quad (10.2)$$

$$\frac{dt}{d\tau} = \left(1 - \frac{u^2}{c^2}\right)^{-\frac{1}{2}} = \xi \quad (10.3)$$

$$U = \frac{dx^\mu}{d\tau} \quad (10.4)$$

$$A = \frac{\partial^2 x^\mu}{\partial \tau} \quad (10.5)$$

$$U = \xi(c, \vec{u}) \quad (10.6)$$

$$A = \xi \frac{d}{dt}(\xi c, \xi u) \quad (10.7)$$

## 11. Relativistic Particle mechanics (in Ω)

The layout[#] [1] here also remains unaffected as shown under (due to same reason as in section 10)

$$P = m_0 U = m_0 \xi(c, u) = (mc, p) \quad (11.1)$$
$$m_0^2 c^4 = m^2 c^4 - p^2 \quad (11.2)$$
$$m = \xi m_0 \quad (11.3)$$
$$E = mc^2 \quad (11.4)$$

## 12. Quantum Mechanics[#] (in Ω)

Here also from the above description it is clear that the structure will remain untouched because of the same reasons as in section 10. Bell's remarks [33] also strongly support this notion.

## 13. Do Lorentz transformations equations play any role for super-luminal frames? (SPECULATIVE)

For this one can consider the aspects of length contraction and time dilation as they directly depends on these.

We have, (note that here L and t represents change in quantities)

$$L = \frac{L_0}{\xi} \quad (13.1)$$

$$t = \xi t_0 \quad (13.2)$$

Multiplying with −1 on the numerator and denominator of (13.1) and bringing −1 in denominator under square root we get

$$L = \frac{-L_0}{\xi} \quad (13.3)$$

Let $m_v$ be the mass of any tachyon, then multiplying with –1 on the numerator and denominator of (13.3) and converting $m_v$ in denominator we get

$$L = \frac{-m_v L_0}{\xi m_0 \xi} \quad (13.4)$$

Then following the similar mode of approach in another deduction[34]; and later extending to solve the imaginary number $\xi$ and then cutting the mass related quantities in the equation we get,

$$L = \frac{-L_0}{\sigma} \quad (13.5)$$

Where $\sigma$ is the positive quantity obtained[34] equivalent of $\xi$ from the same deduction[34]
Similarly we can have,

$$t = -\sigma t_0 \quad (13.6)$$

**13.a Interpretation**

(13.5) and (13.6) are the ones which shows the property of back ward time travel. And more over when one tries to get value for velocity, it is positive quantity and is in accordance with our assumption. Even in a frame moving like this wrt Ω, this is the case with.

*The author would like to thank Dr. Maurizio Consoli of Italian National Institute of Nuclear Physics for the discussions in the matter and by pointing out errors in the manuscript and style of presentation.*

# IV. References

[1] A.Einstean, Ann. d. Phys.17, 891

[2] J. W. Moffat, Int. J. Mod. Phys. D 2, 351 (1993).

[3] A. Albrecht and J. Magueijo, Phys. Rev. D 59, 043516 (1999).

[4] J.D. Barrow, Phys. Rev. D 59, 043515 (1999).

[5] J.D. Barrow and J. Magueijo. Phys. Lett. B. 447,246(1999).

[6] J.D. Barrow and J. Magueijo, Phys. Lett. B. 443, 104 (1998).

[7] J.D. Barrow and J. Magueijo, Class. Quant. Grav. 16, 1435, 1999.

[8] J. Magueijo, Phys. Rev. D 62, 103521, (2000)

[9] A. Peres, gr-qc 0210066; Phys. Rev. Lett. 19 (1967) 1293

[10] H.A. Lorentz, "Electric phenomena in a system moving with any velocity less than that of light", in The principle of relativity (Collection of original papers on relativity), Dover, New-York, 1952.
[10.a] G. F. Fitzgerald, Science 13, 390 (1889).

[11] L. D. Landau and E. M. Lifshitz, The Classical Theory of Fields (Pergamon Press, Oxford, 1975).

[12] *The Astrophysical Journal*, 429: 429-491

[13] *The American Journal of Science*, No. 203 Vol. XXXIV

[14] Paper by William Hicks as quoted in *New Scientist* Vol 186 No 2493 p 30

[15] M. Consoli and E. Costanzo, *From classic to modern ether-drift experiments: thenarrow window for a preferred frame*, Phys. Lett. A **333** (2004) 355.

[16] *Astrophysics and Space Science* 139 (1987) 389-411

[17] B. Nodland, J. P. Ralston: "Indication of Anisotropy in Electromagnetic Propagation over Cosmological Distances", *Physical Review Letters* 78 (1997), No. 16. 3043-3046

[18] M. Alcubierre: "The warp drive: hyper-fast travel within general relativity". *Classical and Quantum Gravity* 11 (1994), pgs. L73-L77)

[19] "A Physical theory based solely on the first postulate of relativity" P*hysics Letters A* 196 (1994) 1-6

[20] Allias, Aero/Space Engineering, vol 9, p 46

[21] Saxl and Allen – Physical Review D, vol 3, p 823

[22] T. Jacobson and Mattingly, Phys. Rev. D 64, 024028 (2001).

[23] A. Albrecht and J. Magueijo, Phys. Rev. D 59, 043516 (1999).

[24] A. Guth, Phys.Rev. D23 347 (1981); A. Linde, Phys. Lett B108, 1220 (1982);

[25] A.Albrecht and P. Steinhardt, Phys.Rev.Lett. 48 1220 (1982); A. Linde, Phys. Lett B 129, 177 (1983).

[26] J. Magueijo and K. Baskerville, astro-ph/9905393, published by CUP, 2000.

[27] P. Avelino and C. Martins, Phys. Rev. D67, 027302, 2003.

[28] H. Brown, *Einstein's misgivings about his 1905 formulation of special relativity*, Eur. J. Phys. **26** (2005) S85.

[29] Referred in *"Introduction to Special Relativity"* - Oxford University Press

[30] P. Beckmann, Einstein plus two. The Golem press, Boulder, Co (1987).

[31] D. Mitsopoulos, Phys. Essays. 6, 233, (1993).

[32] V.I. Makarov, L'aberration astronomique, private communication.

[33] Bell J S, Speakable and Unspeakable in Quantum Mechanics, Cambridge University Press, 1988

[34] Aasis Vinayak PG, On a heuristic point regarding the origin of tachyons and a revamped definition for them, physics/0511253